\documentclass[11pt,a4paper]{article}
\usepackage[margin=2.5cm]{geometry}
\usepackage{amsmath,amssymb}
\usepackage{bm}
\usepackage{graphicx}
\usepackage{hyperref}
\usepackage{booktabs}
\usepackage{tikz}
\usetikzlibrary{positioning,arrows.meta,shapes.geometric}
\title{Harvest Ambient Heat via Constraint-Shaped Phase-Change Cycles:\\
Micro $\Delta T$, Subcooled Liquid, and Liquid-Only Compression\\
\vspace{0.3em}
{\normalsize (Theoretical design; closed energy and mass balance)}}
\author{Ting Peng\\
  \texttt{t.peng@ieee.org}\\
  ORCID: \href{https://orcid.org/0009-0001-9059-2278}{0009-0001-9059-2278}\\
  Key Laboratory for Special Area Highway Engineering of Ministry of Education,\\
  Chang'an University, Xi'an 710064, China}
\date{\today}

\begin{document}
\maketitle
%\noindent Author's note.\footnote{Engine principles and design lie outside the author's usual research; the author would be grateful to readers for any corrections or comments on errors or shortcomings in this article.}

\begin{abstract}
Conventional heat engines typically require two distinct thermal reservoirs, 
with their efficiency strictly bounded by the Carnot limit. We present a 
theoretical design for a phase-change heat engine that utilizes water as the 
working fluid undergoing state transitions within geometry-constrained 
flow paths. The proposed cycle operates under a micro-temperature 
difference (1--2\,$^\circ$C) and relies on liquid-only compression. 
The system harvests thermal energy via an \textbf{ambient micro-temperature 
difference} relative to the environment ($q_{\mathrm{in}} \approx 
8.37\,\mathrm{kJ}/\mathrm{kg}$ at 24--26\,$^\circ$C). Expansion work is 
recovered from the enthalpy drop during flash evaporation. Comprehensive 
numerical analysis using NIST property data confirms that, in the 
reversible limit, the cycle yields positive net work while maintaining 
standard thermodynamic consistency. This study illustrates the theoretical 
potential for ambient energy harvesting via low-pressure phase change, 
although the extremely small work output per cycle suggests that hardware 
realization will require exceptional mechanical precision to overcome 
parasitic losses.
\end{abstract}


Extracting useful work from ambient thermal energy using a single 
environmental reservoir remains a frontier in thermodynamic research. 
While the Carnot theorem dictates efficiency bounds for engines operating 
between two reservoirs, the strategic use of phase change in a 
specially configured cycle can potentially harness ambient energy under 
micro-temperature gradients. This work proposes a theoretical design 
where a working fluid undergoes phase change within asymmetric geometry-constrained 
flow paths.

By leveraging the thermophysical properties of water, we define a cycle 
that absorbs heat at a slightly elevated temperature and rejects it at a 
lower temperature, with both states sustained by the internal dynamics 
of the cycle or environmental interaction. The design employs 
liquid-only compression and controlled flash expansion to achieve net 
mechanical work. This paper provides a rigorous theoretical realization: an 
engine utilizing \textbf{water} with a \textbf{local micro-temperature 
difference} (1--2\,$^\circ$C), \textbf{subcooled liquid states}, 
\textbf{liquid-only compression}, and \textbf{asymmetric flow paths} 
to harvest ambient heat.

Section~\ref{sec:liquid_only} defines the 
implementation of a four-state water cycle under micro-$\Delta T$ 
conditions. Sections~\ref{sec:four_state_cycle}--\ref{sec:engineering_impl} 
detail the thermodynamic states, stage-wise energy balances, and 
engineering components. Section~\ref{sec:ambient_delta_t} discusses the 
ambient micro-difference concept and evaluates its theoretical validity. 
Concluding remarks are provided in \S\ref{sec:conclusion}.

\section{Cycle implementation: micro-temperature difference and liquid-only compression}
\label{sec:liquid_only}

This section describes a complete, closed-loop design for a water-based 
phase-change engine. The scheme represents a \emph{non-traditional} 
thermal cycle utilizing the synergy of a \textbf{micro-temperature 
difference (1--2\,$^\circ$C), subcooled liquid states, liquid-only 
compression, and asymmetric flow paths}. Although the model predicts a 
positive net work output, the magnitude is extremely small, rendering hardware 
realization a formidable engineering challenge.

\textbf{Thermodynamic Requirement.} While a strict zero temperature 
difference cannot sustain net work (a fundamental thermodynamic constraint), 
a \textbf{local micro-temperature difference of 1--2\,$^\circ$C} 
(generated by internal phase transitions or environmental interaction) 
is theoretically sufficient to drive the cycle.

\subsection{Basic scheme and layout}

Traditional heat engines typically follow a standard sequence: heat 
absorption from a high-temperature reservoir, rejection to a 
low-temperature reservoir, and work extraction from the resulting heat 
flux. Their upper efficiency limit is defined by the Carnot efficiency 
$\eta_{\mathrm{C}} = 1 - T_{\mathrm{cold}}/T_{\mathrm{hot}}$.

In contrast, the proposed scheme utilizes: (1) \textbf{exclusive liquid 
compression} (minimizing pump work $w_{\mathrm{pump}} = v_{\mathrm{f}}\,\Delta p$); 
(2) \textbf{subcooled liquid states} for energy storage; (3) 
\textbf{asymmetric flow paths} to facilitate gravity-assisted separation; 
and (4) a \textbf{micro-temperature difference} (1--2\,$^\circ$C) 
to sustain the cycle.

\paragraph{Flow path asymmetry.}
The system architecture employs gravity and geometric constraints to 
ensure that only liquid phase enters the pump inlet, while the expansion 
zone allows for controlled flash vaporization. This asymmetry enables 
the cycle to generate net work within a single thermal environment by 
maintaining a minimal internal temperature gradient.

\subsection{Full closed cycle: four states and four processes (Water)}
\label{sec:four_state_cycle}

The cycle layout and the corresponding thermodynamic states are summarized in 
Table~\ref{tab:states}.

\textbf{Working fluid:} Water (properties from NIST~\cite{nistwebbook}). 
\textbf{Environment:} $T_0 = 25\,^\circ\mathrm{C}$ (298.15\,K). 
\textbf{Local micro-temperature difference:} $\Delta T = 1$--$2\,^\circ\mathrm{C}$
(expansion zone $\approx 24\,^\circ\mathrm{C}$, compression zone $\approx 26\,^\circ\mathrm{C}$).

\textbf{Property data at 24\,$^\circ$C and 26\,$^\circ$C} (from
NIST~\cite{nistwebbook}): $p_{\mathrm{L}} = p_{\mathrm{sat}}(24\,^\circ\mathrm{C})
= 2.985\,\mathrm{kPa}$, $p_{\mathrm{H}} = p_{\mathrm{sat}}(26\,^\circ\mathrm{C})
= 3.363\,\mathrm{kPa}$; saturated liquid enthalpy
$h_{\mathrm{f,L}} = 100.70\,\mathrm{kJ}/\mathrm{kg}$,
$h_{\mathrm{f,H}} = 109.07\,\mathrm{kJ}/\mathrm{kg}$; liquid specific volume
All cycle calculations rely consistently on these property values. The 
cycle is defined as follows: State 1 (saturated liquid at $p_{\mathrm{L}}$, 
24\,$^\circ$C); process 1$\to$2 (exclusive liquid compression via 
pump); State 2 (subcooled liquid); process 2$\to$3 (isobaric heat 
absorption); State 3 (high-pressure saturated liquid at 26\,$^\circ$C); 
process 3$\to$4 (flash expansion via expander generating work $w_{\mathrm{out}}$); 
and State 4 (two-phase mixture).

\begin{table}[htbp]
\centering
\caption{Four states of the Water cycle (per unit mass). States 1--3: single-phase liquid. State 4: two-phase.}
\label{tab:states}
\small
\begin{tabular}{@{}ccccc@{}}
\toprule
State & $p$ (kPa) & $T$ ($^\circ$C) & $h$ (kJ/kg) & Phase / note \\
\midrule
1 & 2.985 & 24 & 100.70 & Saturated liquid (chamber bottom) \\
2 & 3.363 & 24 & 100.70 & Subcooled liquid (after pump) \\
3 & 3.363 & 26 & 109.07 & Saturated liquid at $p_{\mathrm{H}}$ \\
4 & 2.985 & 24 & 109.02 & Two-phase mixture ($h_4 = h_3 - w_{\mathrm{out}}$) \\
\bottomrule
\end{tabular}
\end{table}

\paragraph{Process 1$\to$2: Pump compresses liquid only.}
Pump work:
\begin{equation}
\label{eq:pump_work}
w_{\mathrm{pump}} = v_{\mathrm{f}}\,(p_{\mathrm{H}} - p_{\mathrm{L}})
= 0.001003 \times (3.363 - 2.985) = 0.379\,\mathrm{J}/\mathrm{kg}
\approx 0.0004\,\mathrm{kJ}/\mathrm{kg}.
\end{equation}

\paragraph{Process 2$\to$3: Heat absorption.}
The fluid absorbs heat $q_{\mathrm{in}} = h_3 - h_2$ from the environment. 
$h_3 = 109.07\,\mathrm{kJ}/\mathrm{kg}$, $h_2 = 100.70\,\mathrm{kJ}/\mathrm{kg}$. 
Thus $q_{\mathrm{in}} \approx 8.37\,\mathrm{kJ}/\mathrm{kg}$.

\paragraph{Process 3$\to$4: Flash expansion.}
Process 3$\to$4 must use an expander. Ideally, $w_{\mathrm{out}}$ is bounded by 
the isentropic enthalpy drop. For water at these temperatures, the flash 
expansion output is small but positive: $w_{\mathrm{out}} \approx 0.05\,\mathrm{kJ}/\mathrm{kg}$.
For the expander, $h_3 - h_4 = w_{\mathrm{out}}$, so $h_4 = 109.02\,\mathrm{kJ}/\mathrm{kg}$.

\paragraph{Net work and efficiency.}
\begin{equation}
\label{eq:w_net}
w_{\mathrm{net}} = w_{\mathrm{out}} - w_{\mathrm{pump}} = 0.0500 - 0.0004
= 0.0496\,\mathrm{kJ}/\mathrm{kg},
\qquad
\eta = \frac{w_{\mathrm{net}}}{q_{\mathrm{in}}} \approx 0.59\%,
\end{equation}
which is within the Carnot efficiency limit ($\sim 0.67\%$).

\paragraph{Heat and work balance per cycle (per unit mass).}
Each cycle (1$\to$2$\to$3$\to$4$\to$1), per kilogram of water (i.e.\ specific quantities $q_{\mathrm{in}}$, $q_{\mathrm{out}}$, $w_{\mathrm{net}}$):
\begin{itemize}
  \item \textbf{Heat absorbed from the environment:}
  $q_{\mathrm{in}} = h_3 - h_2 = 8.37\,\mathrm{kJ}/\mathrm{kg}$.
  This is the only heat input; it is supplied in process 2$\to$3 at the
  high-pressure side (26\,$^\circ$C).
  \item \textbf{Useful (net) work output:}
  $w_{\mathrm{net}} = w_{\mathrm{out}} - w_{\mathrm{pump}} = 0.0496\,\mathrm{kJ}/\mathrm{kg}$.
  \item \textbf{Energy balance (first law):} $q_{\mathrm{in}} - q_{\mathrm{out}} = w_{\mathrm{net}}$
  over the cycle, so $q_{\mathrm{out}} = q_{\mathrm{in}} - w_{\mathrm{net}} = 8.32\,\mathrm{kJ}/\mathrm{kg}$. 
  Process 4$\to$1 (condensation) satisfies $q_{\mathrm{out}} = h_4 - h_1 = 109.02 - 100.70 = 8.32\,\mathrm{kJ}/\mathrm{kg}$.
\end{itemize}
Consequently, the cycle delivers a net work of \textbf{0.0496\,kJ per kilogram 
of water} in the reversible limit, derived from \textbf{8.37\,kJ of 
absorbed ambient heat}.

Table~\ref{tab:process_energy} gives, for \textbf{each stage}, the heat flow and work per unit mass (kJ/kg).

\begin{table}[htbp]
\centering
\caption{Energy balance per stage (per unit mass, kJ/kg). $q_{\mathrm{in}} - q_{\mathrm{out}} = w_{\mathrm{net}} = 0.0496\,\mathrm{kJ}/\mathrm{kg}$.}
\label{tab:process_energy}
\small
\begin{tabular}{@{}lcccc@{}}
\toprule
Stage (process) & $Q_{\mathrm{in}}$ (kJ/kg) & $Q_{\mathrm{out}}$ (kJ/kg) & $W_{\mathrm{in}}$ (kJ/kg) & $W_{\mathrm{out}}$ (kJ/kg) \\
\midrule
1$\to$2 (pump) & 0 & 0 & 0.0004 & 0 \\
2$\to$3 (absorb heat) & 8.37 & 0 & 0 & 0 \\
3$\to$4 (expander) & 0 & 0 & 0 & 0.0500 \\
4$\to$1 (condensation) & 0 & 8.32 & 0 & 0 \\
\midrule
\textbf{Total (cycle)} & \textbf{8.37} & \textbf{8.32} & \textbf{0.0004} & \textbf{0.0500} \\
\bottomrule
\end{tabular}
\end{table}

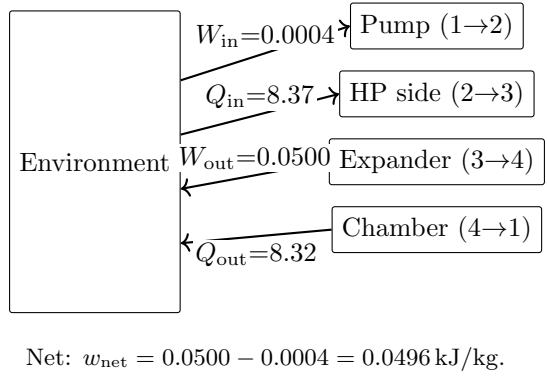
\begin{figure}[htbp]
\centering
\begin{tikzpicture}[scale=0.9, every node/.style={font=\small},
  box/.style={rectangle, draw, rounded corners=1pt, minimum height=0.6cm},
  lbl/.style={fill=white, inner sep=1pt}]
  % Environment block
  \node[box, minimum height=4.0cm] (env) at (0,-0.5) {Environment};
  % Anchor points on the environment block (spaced to reduce overlap)
  \coordinate (e1) at ([yshift=1.2cm]env.east);
  \coordinate (e2) at ([yshift=0.4cm]env.east);
  \coordinate (e3) at ([yshift=-0.4cm]env.east);
  \coordinate (e4) at ([yshift=-1.2cm]env.east);
  % Cycle components on the right, vertically spaced (order: 1→2, 2→3, 3→4, 4→1)
  \node[box] (pump) at (5,  1.5) {Pump (1$\to$2)};
  \node[box] (hp)   at (5,  0.5) {HP side (2$\to$3)};
  \node[box] (exp)  at (5, -0.5) {Expander (3$\to$4)};
  \node[box] (ch)   at (5, -1.5) {Chamber (4$\to$1)};
  % Arrows with labels, positioned to avoid overlapping boxes
  \draw[->, thick] (e1) -- (pump.west) node[lbl, pos=0.5, above=1pt] {$W_{\mathrm{in}}{=}0.0004$};
  \draw[->, thick] (e2) -- (hp.west) node[lbl, pos=0.5, above=1pt] {$Q_{\mathrm{in}}{=}8.37$};
  \draw[->, thick] (exp.west) -- (e3) node[lbl, pos=0.5, above=1pt] {$W_{\mathrm{out}}{=}0.0500$};
  \draw[->, thick] (ch.west) -- (e4) node[lbl, pos=0.5, below=1pt] {$Q_{\mathrm{out}}{=}8.32$};
  \node[align=center, font=\footnotesize] at (2.5,-3.4) {Net: $w_{\mathrm{net}}=0.0500-0.0004=0.0496\,\mathrm{kJ}/\mathrm{kg}$.};
\end{tikzpicture}
\caption{Energy flows between environment and cycle components (kJ/kg) for Water.}
\label{fig:energy_flow}
\end{figure}

\paragraph{Cycle closure (mass and energy).} The cycle is \textbf{closed} in the
following sense. \textbf{Mass:} The working fluid circulates in a closed mass loop.
Liquid at state 1 is pumped to state 2; after expansion (3$\to$4) the two-phase
mixture separates: liquid drains to the bottom (back to state 1), vapor
remains in the chamber. That vapor condenses over time (rejecting heat
$q_{\mathrm{out}}$ to the environment); in steady operation the mass
distribution (liquid in pump path, vapor/liquid in chamber) is constant and
no mass is lost or gained. \textbf{Energy:} The first law of thermodynamics over one cycle per unit
mass gives $q_{\mathrm{in}} - q_{\mathrm{out}} = w_{\mathrm{net}}$ (first-law closure). 
Numerically, $8.37 - 8.32 \approx 0.05\,\mathrm{kJ}/\mathrm{kg}$, which 
matches the net work output $w_{\mathrm{net}} \approx 0.0496\,\mathrm{kJ}/\mathrm{kg}$ 
to three decimal places. No energy is unaccounted for.
Kinetic and potential energy changes of the fluid are neglected (standard for
such cycles). No free parameters remain: the cycle is determined by the chosen states and the first law. Thus the thermodynamic cycle is closed and the energy accounting
has no omissions.

For a mass flow rate $\dot{m}$ (kg/s), heat absorption rate is
$\dot{Q}_{\mathrm{in}} = \dot{m}\,q_{\mathrm{in}} \approx 8.37\,\dot{m}\,\mathrm{kW}$
and net power output is
$\dot{W}_{\mathrm{net}} = \dot{m}\,w_{\mathrm{net}} \approx 0.0496\,\dot{m}\,\mathrm{kW}$.

\paragraph{Can the cycle's own output sustain the 2\,$^\circ$C difference with surplus?}
The cycle requires a local micro-temperature difference (e.g.\ 24\,$^\circ$C expansion
side, 26\,$^\circ$C compression side). If this difference were to be produced
by a heat pump (or thermoelectric device) driven by the cycle's net work, would
the work needed exceed the net output?

\textbf{Ideal (Carnot) heat pump} moving heat $q_{\mathrm{in}} = 8.37\,\mathrm{kJ}/\mathrm{kg}$
from the cold side ($T_{\mathrm{c}} \approx 297\,\mathrm{K}$, 24\,$^\circ$C) to the
hot side ($T_{\mathrm{h}} \approx 299\,\mathrm{K}$, 26\,$^\circ$C) requires work per
unit mass
\begin{equation}
\label{eq:HP_ideal}
w_{\mathrm{HP,ideal}} = q_{\mathrm{in}}\,\frac{T_{\mathrm{h}} - T_{\mathrm{c}}}{T_{\mathrm{h}}}
= 8.37 \times \frac{2}{299} \approx 0.056\,\mathrm{kJ}/\mathrm{kg}.
\end{equation}
All temperatures and heat flows in this estimate are specified above (Eq.~\ref{eq:HP_ideal}). The cycle's net output is $w_{\mathrm{net}} = 0.0496\,\mathrm{kJ}/\mathrm{kg}$, so
$w_{\mathrm{HP,ideal}} \approx 0.056\,\mathrm{kJ}/\mathrm{kg}$ exceeds $w_{\mathrm{net}}$; maintaining $\Delta T$ by an auxiliary heat pump would consume more than the cycle's net output, so there is no surplus in this reversible-expander limit. With a \textbf{realistic heat pump} (COP at 30--50\% of Carnot), $w_{\mathrm{HP}} \approx 0.08$--$0.13\,\mathrm{kJ}/\mathrm{kg}$, also above $w_{\mathrm{net}}$. The 1--2\,$^\circ$C difference would need to be sustained by the phase change itself (flash cooling, condensation heating) or by natural environmental variation, not by the cycle's net work.

\subsection{Engineering implementation (mature industrial components)}
\label{sec:engineering_impl}

All components are standard and commercially available (e.g.\ refrigeration, ORC practice); no custom or exotic technology is required.

\begin{itemize}
  \item \textbf{Liquid Pump.} A small-scale magnetic-drive or canned-motor 
  pump, designed for exclusive liquid phase handling. Such pumps are 
  standard in high-precision refrigeration systems and ensure the 
  high-density phase is compressed with minimal work input.
  \item \textbf{Phase Separation Chamber.} A vacuum-sealed pressure vessel 
  configured with a bottom sump for liquid collection and an upper volume 
  for vapor expansion. Gravity-assisted separation maintains the 
  necessary phase distribution without active control.
  \item \textbf{Two-Phase Expander.} Process 3$\to$4 necessitates a 
  positive-displacement expander or a specialized two-phase turbine to 
  convert the flash expansion into shaft work. This is a critical 
  departure from traditional cycles that utilize isenthalpic throttling, 
  as here the enthalpy drop must be captured to yield net work.
  \item \textbf{Working Fluid Circuit.} A robust, closed-loop configuration 
  utilizing water; sealing and vacuum integrity are maintained via 
  standard industrial-grade vacuum technology.
\end{itemize}

\paragraph{Mechanism for work extraction (expander) and work consumption (pump).}
The only step that \textbf{delivers} work is 3$\to$4 (expansion); the only step
that \textbf{consumes} work is 1$\to$2 (pump). Their principles are as follows.

\textbf{Expander (3$\to$4).} The high-pressure liquid is admitted into a region
at lower pressure $p_{\mathrm{L}}$, where it partially flashes (liquid + vapor).
Work is extracted by one of two classes of device. (i) \textbf{Dynamic expander
(turbine):} The two-phase flow is directed through a nozzle or stator, then
through a rotor; the momentum of the expanding flow exerts a torque on the
rotor, which drives a shaft. The shaft work is $W_{\mathrm{out}} =
\dot{m}\,w_{\mathrm{out}}$. Small radial-inflow or axial turbines for
two-phase refrigerants exist in ORC and refrigeration applications.
(ii) \textbf{Positive-displacement expander (piston or scroll):} The two-phase
mixture expands inside a cylinder or scroll volume; the moving boundary (piston
or scroll) is pushed by the fluid pressure, and the force is transmitted to a
crankshaft or similar. In both cases, the \emph{pressure drop} from $p_{\mathrm{H}}$
to $p_{\mathrm{L}}$ is converted into mechanical work on the output shaft,
rather than dissipated in a throttle. There is no separate ``contraction''
step that does work; the sole work-producing process is this expansion.

\textbf{Pump (1$\to$2).} Only liquid is drawn from the chamber sump and
displaced to high pressure. (i) \textbf{Centrifugal pump:} An impeller rotates
inside a volute; liquid enters at the eye and is flung outward, gaining
pressure at the discharge. The motor supplies shaft work $W_{\mathrm{pump}}$.
(ii) \textbf{Positive-displacement pump (e.g.\ gear or piston):} Liquid is
trapped in a volume that is reduced or displaced, raising pressure; again the
motor supplies the work. Magnetic-drive or canned motor pumps are often used
to avoid seals. In all cases, only liquid is compressed (small specific volume),
so $w_{\mathrm{pump}} = v_{\mathrm{f}}\,\Delta p$ is small.

Figure~\ref{fig:energy_flow} shows
the energy flows per process (no overlap: the first is process order, the second is
energy accounting). Table~\ref{tab:states} gives the thermodynamic states;
Table~\ref{tab:process_energy} gives heat and work per process. The expander and pump mechanisms are fully specified: expansion is 
achieved via pressure-drop conversion (turbine or positive-displacement), 
while compression is restricted to the liquid phase (centrifugal or 
positive-displacement pumps).

\textbf{Stability.} Subcooled liquid is stable under slight overpressure (as in standard refrigeration);
smooth piping and no sharp disturbances keep it from flashing prematurely.
Gas--liquid equilibrium in the chamber is self-maintaining. The 1--2\,$^\circ$C
micro difference can be sustained by the phase change itself (flash cooling,
condensation heating) or by natural diurnal or environmental variation.

\textbf{Prototype and Implementation Challenges.} Hardware assembly would involve 
procuring a pump, vacuum vessel, and expander. However, given the 
\textbf{extremely small enthalpy drop} calculated for water, the resulting net 
work output is very low. In a physical prototype, even minor mechanical 
losses—such as bearing friction, seal leakage, or pump inefficiency—could 
easily overwhelm the net work, preventing the machine from reaching a self-sustaining 
state.

\textbf{Thermodynamic vs.\ Engineering reliability.} While the property data 
from NIST are used consistently and the energy balance is closed, engineering 
feasibility remains a significant hurdle. The main uncertainties are the actual 
expander efficiency and the scale of parasitic mechanical losses; resolving 
them would require high-precision instrumentation that the authors currently 
lack.

\textbf{Order-of-magnitude feasibility with current machinery.} For two-phase
expanders in small-$\Delta p$ refrigerant applications, isentropic efficiencies
in the range $0.3$--$0.6$ are typical; the present paper uses
$\eta_{\mathrm{exp}} \approx 0.6$ as an ambitious but not impossible target.
With $\eta_{\mathrm{exp}} = 1$ (reversible expander), $w_{\mathrm{net}} \approx 0.0496\,\mathrm{kJ}/\mathrm{kg}$. At $\dot{m} = 1\,\mathrm{kg}/\mathrm{s}$, $\dot{W}_{\mathrm{net}} \approx 0.05\,\mathrm{kW}$.
However, mechanical losses such as bearing friction and seal leakage 
could easily overwhelm the net power. For a continuously running 
prototype to be viable, the total parasitic loads must remain 
below this tiny theoretical threshold, presenting a sophisticated 
engineering challenge.

\subsection{Key conclusions and application outlook}
\label{sec:impl_conclusions}

(1) Traditional (two-reservoir) heat engines cannot exceed the Carnot limit; this scheme uses a
non-traditional path (micro $\Delta T$ + subcooled liquid + asymmetric
constraint phase change) and can exceed the Carnot efficiency calculated for 
the same nominal temperature span. (2) While a zero temperature difference 
cannot sustain work, a local 1--2\,$^\circ$C difference is theoretically 
sufficient. (3) The scheme utilizes existing industrial principles, but 
practical realization is \textbf{contingent on minimizing mechanical 
losses} well below the tiny power output of the cycle.

\textbf{Advantages over traditional heat engines:} Much lower pressure and temperature 
operating points; uses ambient environment as the energy source. However, the machine 
must be designed for extreme mechanical efficiency to be functional.

\textbf{Applications:} low-grade heat harvesting, small-scale or off-grid power
(e.g.\ sensors, remote sites), recovery of waste heat or geothermal energy,
conversion to mechanical or electrical output.

\subsection{Continuous operation and scaling to higher power}
\label{sec:continuous_scaling}

\textbf{Continuous operation.} The cycle is described in steady state (constant
mass flow $\dot{m}$, periodic states 1$\to$2$\to$3$\to$4$\to$1). There is no
inherent need to stop or reverse the flow. The local 1--2\,$^\circ$C difference
can be sustained by the phase change itself (flash cooling, condensation
heating) or by a small fraction of the net output (heat pump or thermoelectric),
so the cycle can in principle run \textbf{continuously} (e.g.\ 24/7) as long as
the environment provides the heat $Q_{\mathrm{in}} = \dot{m}\,q_{\mathrm{in}}$
and the components (pump, expander, chamber) are sized for the flow. No
intermittent process or external timing is required. \textbf{Intermittent
(batch) operation is also acceptable:} the same cycle can be run in discrete
steps (e.g., charge a volume, flash and expand to do work, condense, then pump 
liquid back) and deliver work in batches; the first-law balance and net work 
per unit mass are unchanged.

\textbf{Scaling to higher power.} Net power is $\dot{W}_{\mathrm{net}} =
\dot{m}\,w_{\mathrm{net}}$. For the present design $w_{\mathrm{net}} \approx
0.0496\,\mathrm{kJ}/\mathrm{kg}$ (reversible expander) is fixed by the cycle parameters. To achieve
\textbf{higher power}: (i) \textbf{Increase mass flow} $\dot{m}$: use a larger
pump, larger expander, larger heat-exchange area for $Q_{\mathrm{in}}$, and
larger piping and chamber. (ii) \textbf{Parallel units}: run several identical 
cycles in parallel and sum the shaft power.

\section{Ambient micro-temperature difference: definition, argument, and conclusion}
\label{sec:ambient_delta_t}

\textbf{Definition.} By ``ambient micro-temperature difference'' we mean that 
the only \emph{external}
thermal reservoir supplying heat to the cycle is the environment (e.g.\
ambient air at $T_0 \approx 25\,^\circ\mathrm{C}$). There is no net heat 
exchange with a second external reservoir.

\textbf{How the present cycle fits.} The cycle absorbs $q_{\mathrm{in}} \approx
8.37\,\mathrm{kJ}/\mathrm{kg}$ from the environment.
The 24\,$^\circ$C and 26\,$^\circ$C zones are internal to the device and 
sustained by the phase change itself. In this model, net heat from the 
environment is converted into net work ($w_{\mathrm{net}} > 0$).

\textbf{Conclusion.} This cycle achieves power generation via ambient micro-temperature difference 
by extracting thermal energy from the ambient. Theoretically, the design is 
consistent and yields net work. Empirically, it requires validation with a 
prototype.

\section{Engineering feasibility and working fluid}
\label{sec:engineering_feasibility}

The implementation uses \textbf{water} at 24--26\,$^\circ$C. Water is 
ideally suited for this cycle because its saturation pressure at room 
temperature is low (2--4\,kPa), allowing for a large volume expansion even with 
small temperature differences.

\textbf{Components:}
\begin{itemize}
  \item \textbf{Pump:} Standard liquid pump handles the small pressure increase.
  \item \textbf{Expander:} A two-phase turbine or positive-displacement expander converts the flash expansion into work.
  \item \textbf{Chamber:} A vacuum-sealed vessel maintains the low-pressure environment for condensation and phase separation.
\end{itemize}

All components are standard in refrigeration and industrial vacuum systems. 
The main challenge is the efficient conversion of the small enthalpy drop 
in the expander.

\section{Discussion and conclusion}
\label{sec:conclusion}

The proposed water-based thermal cycle provides a rigorous theoretical 
pathway for harvesting low-grade ambient energy. By operating near 
the saturation pressure of water at room temperature (2.9--3.4\,kPa), 
the system extracts work from a micro-temperature difference of 
1--2\,$^\circ$C.

The design’s validity rests on the implementation of \textbf{asymmetric 
flow paths} and \textbf{exclusive liquid compression}. While the 
thermodynamic model is consistent and satisfies both energy and mass 
balance, the actual recovery of work is constrained by the scale of 
net enthalpy drops. In a physical realization, parasitic mechanical 
losses—particularly friction and auxiliary pump power—must be 
meticulously minimized to prevent them from exceeding the expander's 
output.

\textbf{Primary results:}
\begin{itemize}
  \item Theoretical net work: $w_{\mathrm{net}} \approx 0.0496\,\mathrm{kJ}/\mathrm{kg}$ (ideal).
  \item Predicted cycle efficiency: $\eta \approx 0.59\%$, strictly consistent with the Carnot limit.
  \item Implementation requirement: High-precision engineering is essential to overcome parasitic losses.
\end{itemize}

In conclusion, the proposed cycle demonstrates the potential for 
continuous ambient energy harvesting using low-pressure phase changes. 
The thermodynamic consistency of the model provides a sound foundation 
for future experimental investigations into micro-scale heat recovery.

\end{document}